\documentstyle[sprocl,epsf]{article}
\newcommand{\dd}{{\rm d}}
\newcommand{\hmp}{{hard mass procedure}}
\newcommand{\lmp}{{large momentum procedure}}

\newcommand{\order}[1]{{\cal O}(#1)}

\begin{document}
\title{Expansion Techniques in Massive Quark Production:\\ Results and
  Applications}
\author{J.H. K\"uhn}
\address{Institut f\"{u}r Theoretische Teilchenphysik,\\
  Universit\"{a}t Karlsruhe,
  D--76128 Karlsruhe, Germany}
\date{}
\maketitle
\abstracts{
  Recent progress in the calculation of multi-loop, multi-scale diagrams
  is reviewed. Expansion techniques combined with new developments in
  Computer algebra allow to evaluate the $R$ ratio for massive quarks
  up to order $\alpha_s^2$ and, partly, even $\alpha_s^3$. Similar
  techniques can be applied to Higgs- or $Z$-boson decays and mixed QCD
  and electroweak interactions.
}

\section{Introduction\label{sec::chap1}}
During the past years amazing progress has been made in the experimental
tests of the Standard Model of particle physics. Its electroweak sector
has been scrutinized at LEP, mainly in $Z$ decays and $W$ pair
production, and at the TEVATRON, mainly through the precise measurements
of the $W$ boson and top quark masses. Perturbative QCD has been tested
at electron positron colliders at lower energies as well as in high
energy experiments at LEP, at the TEVATRON through proton-antiproton
collisions, and through lepton-nucleon scattering in particular at HERA.

The large statistics collected in these experiments in conjunction with
the small systematic uncertainty allows to test the theoretical
predictions with high precision, requiring the inclusion of quantum
corrections at least in one-loop approximation, occasionally two or even
three-loop calculations are required. This applies on one hand to
perturbative QCD, for electroweak observables on the other hand either
purely weak two-loop corrections should be included or so-called
``mixed'' QCD and electroweak effects up to two or even three loops.
Considerable progress has been made on the theoretical side which
matches these requirements. Two-loop amplitudes, in particular those
contributing to two-point functions can often be evaluated in closed
form or through straightforward numerical integrals --- even with arbitrary
mass assignments. Three-loop amplitudes, however, are at present only
accessible in a few selected cases. These are, most notably, two-point
functions with massless internal propagators only and vacuum
diagrams with massless and massive propagators --- however, of identical
mass throughout. Other cases of interest are ``on-shell amplitudes''
required for mass renormalization and certain types of integrals
relevant for fermion pair production at threshold.

Despite this seemingly limited set of diagrams a large variety of
problems can be solved with the help of expansion techniques. Frequently
one finds a hierarchy of mass and/or energy scales which suggests to
perform a Taylor series of the integrand, with the resulting integrals
being considerably simpler. However, the integrals are not necessarily
analytic in the expansion parameter, and more refined techniques,
denoted hard mass, large
momentum~\cite{methods,reviews} or threshold
expansion~\cite{BenSmi98,CzaMel98,BenSigSmi98} are required. These can be
formulated in a diagrammatic manner and allow to reduce a given
amplitude into an (infinite) sum of products of simpler ``master''
amplitudes which are known in analytical form. In practice often the
first few terms provide already a sufficiently accurate numerical
answer. In this approach, the demands on computation grow rapidly. Not
only the Dirac algebra and the evaluation of integrals for individual
diagrams has to be performed by computer algebra, with the appearance of
hundreds if not thousands of diagrams also the generation of the basic
diagrams, the application of the hard mass or large momentum procedure,
the transformation of diagrams into algebraic expressions and the
overall book keeping has to be performed automatically. (For a recent
review see~\cite{HarSte:review}.)

In the following talk several characteristic examples will be given,
partly taken from QCD, partly from electroweak interactions. The next
chapter will be concerned with the $\order{\alpha_s^2}$ evaluation of
the cross section for electron positron annihilation into massive
quarks. Information on the vacuum polarization function at small and
large $q^2/(4m^2)$ as obtained via expansion techniques is combined with
the two dominant terms close to threshold. Subsequently a variable
transformation is applied as suggested by the analyticity structure of
$\Pi(q^2)$ in the cut complex plane. The numerical results will be
contrasted with those deduced from the threshold expansion. Techniques,
results and limitations of the large momentum procedure will be
presented in chapter~\ref{sec::chap3}. This includes in particular the
calculation of the vacuum polarization of order $\alpha_s^2$, expanded
in $m^2/q^2$ up to fairly high power. The strategy for an evaluation of
the absorptive part in order $\alpha_s^3 m^2/q^2$ and $\alpha_s^3
(m^2/q^2)^2$ and recent results are given in chapter~\ref{sec::chap4}.
Results of relevance for electroweak measurements are discussed in
chapter~\ref{sec::chap5}. This includes the top quark contribution to
the $\rho$ parameter in three-loop approximation, mixed corrections to
the $Z$ decay rate and ``theory driven'' results for the running QED
coupling at scale $M_Z$. Chapter~\ref{sec::chap6} contains a brief
summary and the conclusions.

\section{Three-Loop Heavy-Quark Vacuum Polarization\label{sec::chap2}}
The measurement of the total cross section for electron positron
annihilation into hadrons allows for a unique test of perturbative QCD.
The decay rate $\Gamma(Z \to \mbox{hadrons})$ provides one of the most
precise determinations of the strong coupling constant $\alpha_s$. In
the high energy limit the quark masses can often be neglected.  In this
approximation QCD corrections to $R \equiv \sigma(e^+ e^- \to
\mbox{hadrons})/ \sigma(e^+ e^- \to \mu^+ \mu^-)$ have been 
calculated~\cite{CheKatTka79DinSap79CelGon80,GorKatLar91}
up to order $\alpha_s^3$.
For precision measurements the dominant mass corrections must be
included through an expansion in $m^2/s$. Terms up to order $\alpha_s^3
m^2/s$ (see~\cite{CheKue90}) and $\alpha_s^2 m^4/s^2$ (see~\cite{CheKue94})
and recently~\cite{rhdiss} even up to $\alpha_s^3 m^4/s^2$ are available
at present, providing an acceptable approximation from the high energy
region down to intermediate energy values.  For a number of
measurements, however, the information on the complete mass dependence
is desirable. This includes charm and bottom meson production above the
resonance region, say above $4.5$~GeV and $12$~GeV, respectively, and,
of course, top quark production at a future electron positron collider.
To order $\alpha_s$ this calculation was performed by
K\"all\'en and Sabry in the context of QED a long time ago~\cite{KaeSab55}.
With measurements of ever increasing precision, predictions
in order $\alpha_s^2$ are needed for a reliable
comparison between theory and experiment. Furthermore, 
when one tries to apply the ${\cal O}(\alpha)$ result 
to QCD, with its running coupling
constant, the choice of scale becomes important.
In fact, the distinction between the two intrinsically different 
scales, the relative momentum versus the center of mass
energy, is crucial for a stable numerical prediction.
This information can be obtained from a full calculation 
to order $\alpha_s^2$ only. 
Such a calculation then allows to predict the cross section 
in the complete energy region where perturbative QCD can be applied
--- from close to threshold up to high energies. It is then only the
region very close to threshold, where the fixed order result remains
inadequate and Coulomb resummation becomes important.
In~\cite{CheKueSte96} results for the cross section were calculated in
order $\alpha_s^2$.  They were obtained from the vacuum polarization
$\Pi(q^2)$ which was calculated up to three loops.  The imaginary part
of the ``fermionic contribution'' --- derived from diagrams with a
massless quark loop inserted in the gluon propagator --- had been
calculated earlier in~\cite{HoaKueTeu95}.  In this latter case all
integrals could be performed to the end and the result was expressed in
terms of polylogarithms.  In~\cite{CheKueSte96} the calculation was
extended to the full set of diagrams relevant for QCD.  Instead of
trying to perform the integrals analytically, information of
$\Pi(q^2)$ from the large $q^2$ behaviour, the expansion around $q^2=0$
and from threshold was incorporated.
\nopagebreak
\subsection[]{Outline of the Calculation~\cite{CheKueSte96}}
The different behaviour at threshold makes it necessary to decompose
$\Pi$ according to its colour structure. It is convenient to
write:
\begin{eqnarray}
\Pi(q^2) &=& \Pi^{(0)}(q^2) 
         + \frac{\alpha_s(\mu^2)}{\pi}C_F\Pi^{(1)}(q^2)
         + \left(\frac{\alpha_s(\mu^2)}{\pi}\right)^2\Pi^{(2)}(q^2)
         + \cdots,
\\
\Pi^{(2)} &=& 
                C_F^2       \Pi_{\mbox{\scriptsize\it A}}^{(2)}
              + C_A C_F     \Pi_{\mbox{\scriptsize\it NA}}^{(2)}
              + C_F T n_l \Pi_{\mbox{\scriptsize\it l}}^{(2)}
              + C_F T     \Pi_{\mbox{\scriptsize\it F}}^{(2)}.
\end{eqnarray}
The same notation is adopted to the
physical observable $R(s)$ which is related to $\Pi(q^2)$ by
\begin{eqnarray}
R(s)   &=&  12\pi\, \mbox{Im}\Pi(q^2=s+i\epsilon).
\end{eqnarray}

The contributions from diagrams with $n_l$ light 
or one massive 
internal fermion loop are denoted
by $C_F T n_l\Pi_l^{(2)}$ and $C_F T \Pi_F^{(2)}$, respectively.
The purely gluonic corrections
are proportional to $C_F^2$ or $C_A C_F$ where the former are the only
contributions in an Abelian theory and the latter are characteristic for
the non-Abelian aspects of QCD.

All steps described below have been performed separately for
the first three contributions to $\Pi^{(2)}$. 
In fact, new information is only
obtained for $\Pi_{\mbox{\scriptsize\it A}}^{(2)}$ 
and $\Pi_{\mbox{\scriptsize\it NA}}^{(2)}$ since
$\mbox{Im}\Pi_{\mbox{\scriptsize\it l}}^{(2)}$ 
is already known analytically~\cite{HoaKueTeu95}.
The contribution from a four-particle cut 
with threshold at $4m$ is given
in terms of a two dimensional integral~\cite{HoaKueTeu95,chkst98}
which can be solved 
easily numerically,
so $\Pi_F^{(2)}$ will not be treated.

Let us now discuss the behaviour of $\Pi(q^2)$ in the three different 
kinematical regions and the approximation method.

{\it Analysis of the high $q^2$ behaviour:}
The high energy behaviour of $\Pi$ provides important 
constraints on the complete answer.
In the limit of small $m^2/q^2$ the constant term and the one
proportional to $m^2/q^2$ (modulated by powers of $\ln(\mu^2/q^2$) have been
calculated a long time ago~\cite{GorKatLar86}. The results for terms up to order $(m^2/q^2)^4$
are described in chapter~\ref{sec::chap3}, provide an important cross
check, however, they are not used for the moment.

{\it Threshold behaviour:}
General arguments based on the influence of Coulomb exchange close to 
threshold, combined with the information on the perturbative QCD
potential and the running of $\alpha_s$ dictate the singularities
and the structure of the leading cuts close to threshold, that
is for small $v=\sqrt{1-4m^2/s}$.
The $C_F^2$ term 
is directly related to the QED result with internal photon lines only.
The leading $1/v$
singularity  and the constant term  of $R_A$
can be predicted from the nonrelativistic
Greens function for the Coulomb potential
and the ${\cal O}(\alpha_s)$ calculation. 
The next-to-leading
term is determined by the combination of one-loop
results again with the Coulomb singularities~\cite{BarGatKoeKun75}. One finds
\begin{eqnarray}
R_{\mbox{\scriptsize\it A}}^{(2)} &=& 
3\left(\frac{\pi^4}{8v} - 3\pi^2 + \ldots\right).
\label{Ra}
\end{eqnarray}

The contributions $\sim C_A C_F$ and $\sim C_F T n_l$ can be treated
in parallel. 
For these colour structures the perturbative QCD potential~\cite{Fis77}
\begin{eqnarray}
 V_{\mbox{\scriptsize QCD}}(\vec{q}\,^2) &=& 
         -4\pi C_F\frac{\alpha_V(\vec{q}\,^2)}{\vec{q}\,^2},
\\
\alpha_V(\vec{q}\,^2)  &=&   \alpha_s(\mu^2)\Bigg[
      1 + \frac{\alpha_s(\mu^2)}{4\pi}\bigg(
          \left(\frac{11}{3}C_A
-\frac{4}{3}T n_l\right)
          \left(-\ln\frac{\vec{q}\,^2}{\mu^2}+\frac{5}{3}\right)
          -\frac{8}{3}C_A            \bigg)         
\label{alphav}
\Bigg]
\nonumber
\end{eqnarray}
will become important. 
The leading $C_AC_F$ and $C_F T n_l$ term in $R$ is proportional
to $\ln v$ and is responsible for the evolution of the
coupling constant close to threshold. Also the constant term can
be predicted by the observation, that the leading term in
order $\alpha_s$ is induced by the potential.
The ${\cal O}(\alpha_s)$ result
\begin{equation}
R=3\frac{v(3-v^2)}{2}
\left(1
       + C_F\frac{\pi^2(1+v^2)}{2v}\frac{\alpha_s}{\pi}+\ldots
     \right)
\end{equation}
is employed to predict the logarithmic and constant $C_FC_A$ and
$C_FTn_l$ terms of ${\cal O}(\alpha_s^2)$ by replacing $\alpha_s$
by $\alpha_V(4\vec{p}\,^2=v^2 s)$ as given in Eq.~(\ref{alphav}).
This implies the following threshold behaviour:
\begin{eqnarray}
R_{\it NA}^{(2)}&=&3\frac{\pi^2}{3}(3-v^2)(1+v^2)
                         \left(
                              -\frac{11}{16}\ln\frac{v^2 s}{\mu^2}
                              +\frac{31}{48}
                     +\ldots    \right),
\label{Rna}
\\
R_l^{(2)}&=&3\frac{\pi^2}{3}(3-v^2)(1+v^2)       
                         \left(   
                                  \frac{1}{4}\ln\frac{v^2 s}{\mu^2}
                                 -\frac{5}{12}
                     +\ldots    \right).
\label{Rnl}
\end{eqnarray}

This ansatz suggested in~\cite{CheKueSte96} can be verified for the $C_F
T n_l$ term where the result is known in analytical form~\cite{HoaKueTeu95} and it has also been confirmed for the NA term where
the leading terms for small $v$ have been derived recently (see below).
Extending the ansatz from the behaviour of the imaginary part close to
the branching point into the complex plane allows to predict the leading
terms of $\Pi(q^2)$ $\sim \ln v$ and $\sim \ln^2 v$.

{\it Behaviour at $q^2=0$:}
Important information is contained in the Taylor series of $\Pi(q^2)$
around zero. The calculation of the first seven nontrivial terms is 
based on the evaluation of three-loop tadpole integrals with 
the help of the algebraic program {\tt MATAD} written in {\tt FORM}~\cite{VerFORM}
which
performs the traces,
calculates the derivatives with respect to the external momenta. 
It reduces the large number of different
integrals to one master integral and a few simple ones
through an elaborate set of
recursion relations based on the integration-by-parts method~\cite{CheTka81,Bro92}.
The result can be written in the form:
\begin{eqnarray}
\Pi^{(2)} &=& 
              \frac{3}{16\pi^2}
              \sum_{n>0} C_{n}^{(2)} \left(\frac{q^2}{4m^2}\right)^n,
\end{eqnarray}
where the first seven moments are listed in~\cite{CheKueSte96}.

\begin{figure}[ht]
 \begin{center}
 \begin{tabular}{c}
   \epsfxsize=7cm
   \leavevmode
   \epsffile[80 280 540 520]{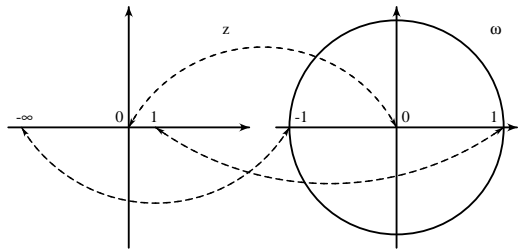}
 \end{tabular}
 \caption{\label{fig::trafo} Transformation between the
                              $z$ and $\omega$ plane }
 \end{center}
\end{figure}

{\it Conformal mapping and Pad\'e approximation:}
The vacuum polarization function 
$\Pi^{(2)}$ is analytic in the complex plane
cut from $q^2=4m^2$ to $+\infty$. The Taylor series in $q^2$ is
thus convergent in the domain $|q^2|<4m^2$ only. The
conformal mapping 
which corresponds to the variable 
transformation ($z=q^2/(4m^2)$)
\begin{eqnarray}
\omega = \frac{1-\sqrt{1-z}}{1+\sqrt{1-z}},
\qquad
z = \frac{4\omega}{(1+\omega)^2},
\label{omega}
\end{eqnarray}
transforms the cut complex $z$ plane into the
interior of the
unit circle. The special points
$z=0,1,-\infty$ correspond to $\omega=0,1,-1$, respectively
(Fig.~\ref{fig::trafo}).

The upper (lower) part of the cut is mapped onto the upper (lower) 
perimeter of the circle.
The Taylor series in $\omega$ thus converges in the interior of the
unit circle. To obtain predictions for
$\Pi(q^2)$ at the boundary it has been suggested~\cite{FleTar94,BroFleTar93}
to use
the Pad\'e approximation 
which converges towards $\Pi(q^2)$ even on the perimeter.
To improve the accuracy 
the singular threshold behaviour  
and the large $q^2$ behaviour
is incorporated into simple analytical functions
which are removed 
from $\Pi^{(2)}$ before the Pad\'e approximation is 
performed.
The quality of this
procedure can be tested by comparing the prediction with
the known result for $\mbox{Im}\Pi_{\mbox{\scriptsize\it l}}^{(2)}$.

The logarithmic singularities at threshold and large $q^2$
are removed by subtraction, the $1/v$ singularity, which is present
for the $C_F^2$ terms only, by multiplication with $v$ as 
suggested in~\cite{BaiBro95}.
The imaginary part of the remainder which is actually 
approximated by the Pad\'e method is thus smooth in the
entire circle, numerically small and vanishes at 
$\omega=1$ and $\omega=-1$. 

\tabcolsep=.4em
\begin{figure}[ht]
 \leavevmode
 \begin{center}
 \begin{tabular}{cc}
   \epsfxsize=5.5cm
   \epsfysize=4cm
   \epsffile[110 330 460 520]{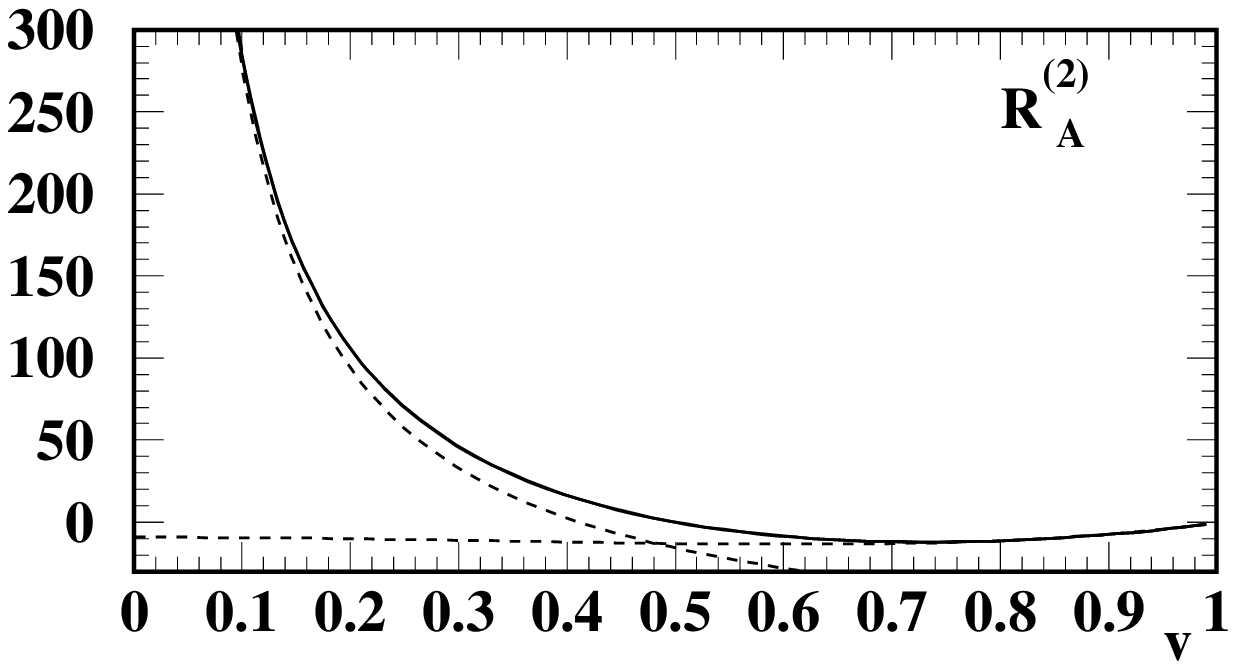}  
   &
   \epsfxsize=5.5cm
   \epsfysize=4cm
   \epsffile[110 330 460 520]{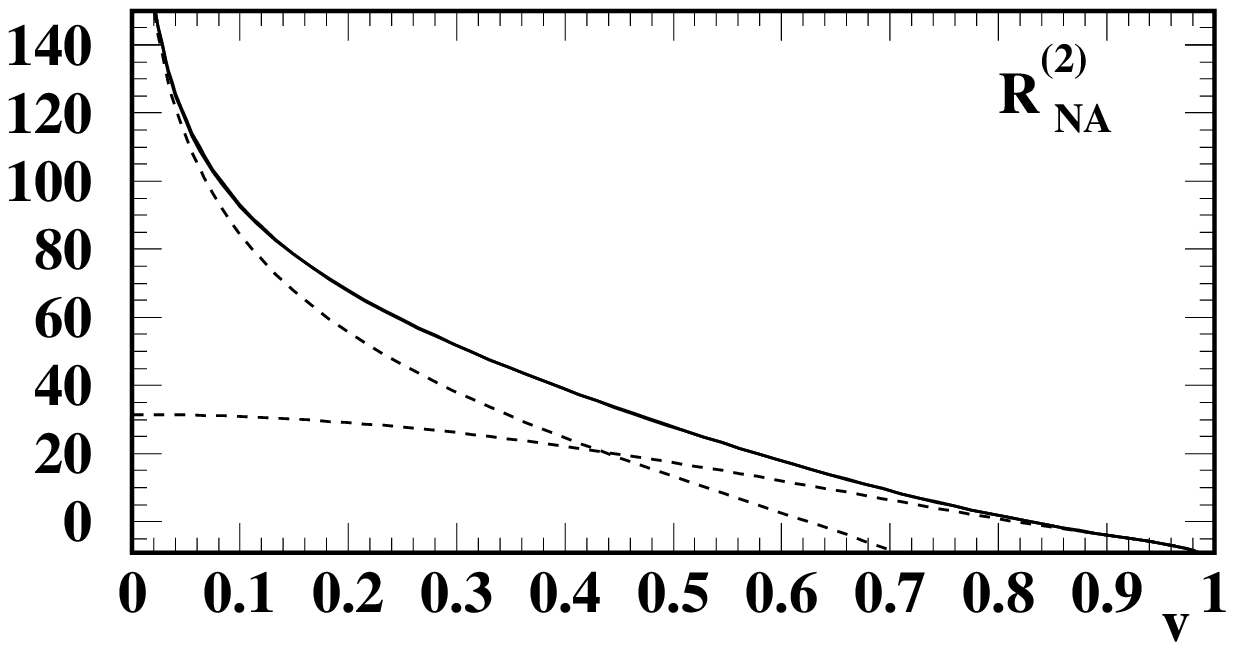} 
 \end{tabular}
 \caption{\label{fighigh}Complete results (full line) are
                         compared to the threshold
                         approximations and the high energy
                         approximations including the $m^2/s$
                         (dash-dotted) and the $m^4/s^2$ (dashed) terms
                         ($x=2m/\protect\sqrt{s}$).
         }
 \end{center}
\end{figure}
\begin{figure}[ht]
 \leavevmode
 \begin{center}
 \begin{tabular}{cc}
   \epsfxsize=5.5cm
   \epsfysize=4cm
   \epsffile[110 330 460 520]{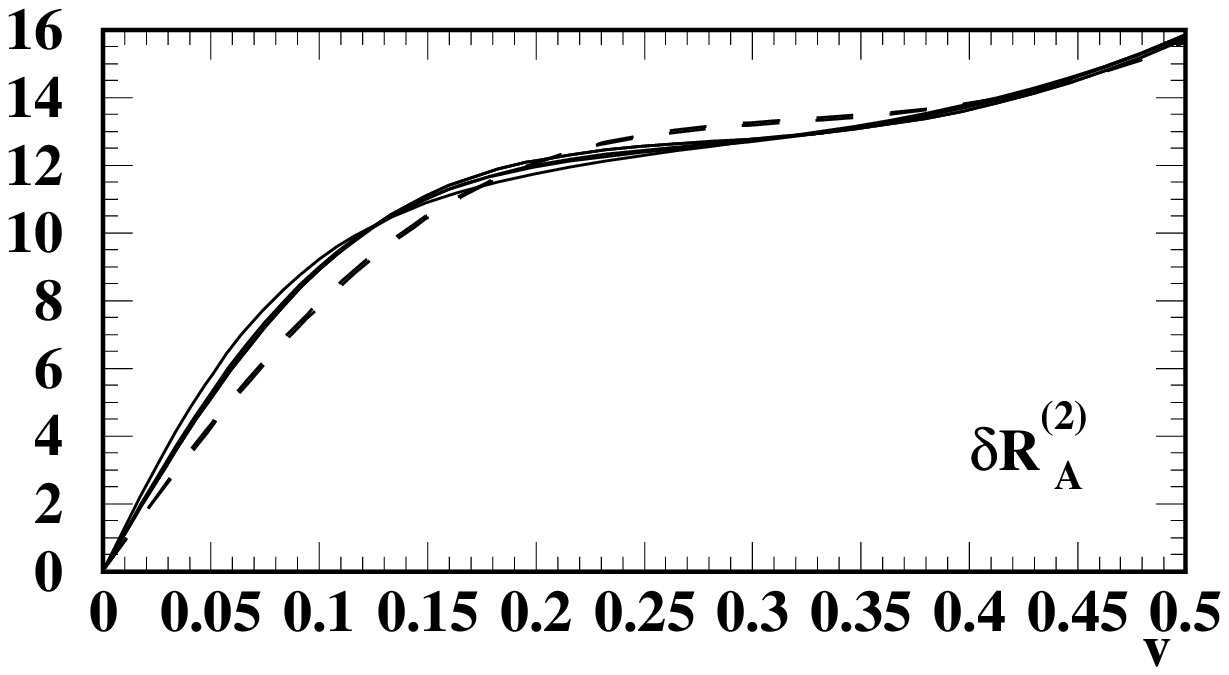}  
   &
   \epsfxsize=5.5cm
   \epsfysize=4cm
   \epsffile[110 330 460 520]{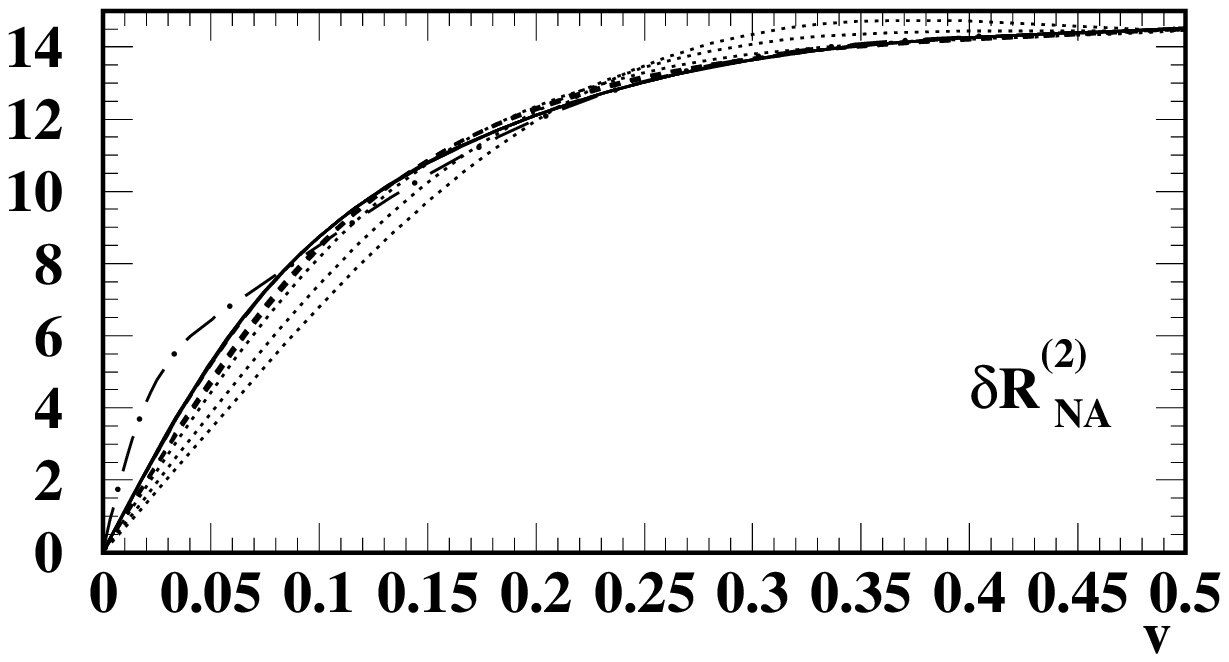} 
 \end{tabular}
 \caption{\label{figthr}The 
                        threshold behaviour of the remainder $\delta R$ 
                        for three different Pad\'e approximants is shown.
                        (The singular
                        and constant parts around threshold are subtracted.)
        }
 \end{center}
\end{figure}
\tabcolsep=0em

\subsection[]{Results}

After performing the Pad\'e approximation for the smooth remainder
with $\omega$ as natural variable, the transformation (\ref{omega})
is inverted and the full vacuum polarization function reconstructed
by reintroducing the threshold and high energy terms. This
procedure provides real and imaginary parts of $\Pi^{(2)}$.
Subsequently only the absorptive part of $\Pi^{(2)}$ (multiplied by
$12\pi$) will be presented. 

In the following it will be useful to plot the results as functions of
$x=2m/\sqrt{s}$ and alternatively of $v=\sqrt{1-4m^2/s}$. The first choice
is particularly useful for investigations of the high energy region, the
second one for energies close to threshold. Characteristic values of
$x$, $v$ and $\sqrt{s}$ for charm ($m_c=1.6$~GeV), bottom ($m_b=4.7$~GeV)
and top ($m_t = 175$~GeV) quarks are listed in Table~\ref{tab::1} for
easy comparisons.
\tabcolsep=.19em
\begin{table}
\begin{tabular}{|l|r|r|r|r|r|r|r|r|r|r|r|r|}
\hline
x & 0.1 & 0.2 & 0.3 & 0.4 & 0.5 & 0.6 & 0.7 & 0.8 & 0.9 & 0.95 & 0.97 & 0.98 \\
\hline
v & 0.995 & 0.980 & 0.954 & 0.917 & 0.866 & 0.800 & 0.714 & 0.600 & 0.436 &
0.312 & 0.243 & 0.199\\
\hline
$\sqrt{s_c}$
& 32 & 16 & 10.7 & 8.0 & 6.4 & 5.3 & 4.6 & 4.0 & --- &--- & --- & --- \\
\hline
$\sqrt{s_b}$
& 94 & 47 & 31.3 & 23.5 & 18.8 & 15.7 & 13.4 & 11.8 & 10.4 & --- & --- &
--- \\
\hline
$\sqrt{s_t}$
& 3500 & 1750 & 1167 & 875 & 700 & 583 & 500 & 438 & 389 & 368 & 361 &
357\\
\hline
\end{tabular}
\caption{\label{tab::1} Conversion between $x$ and $v$ and values for
$\sqrt{s}$ in GeV for charm, bottom and top production.}
\end{table}
\tabcolsep=0em
Energy values where a perturbative treatment is evidently inapplicable
are denoted by dashes.

In Fig.~\ref{fighigh} the complete results are shown 
for $\mu^2=m^2$ with 
$R_{\mbox{\scriptsize\it A}}^{(2)}$ and $R_{\mbox{\scriptsize\it NA}}^{(2)}$
displayed separately. 
The solid line represents the full correction. The threshold
approximation is given  by the dashed curve. In the
high energy region the corrections containing the $m^2/s$ terms and
the quartic approximations are included.
It should be stressed that the latter are not incorporated into the 
construction of $R^{(2)}$ but they are evidently very well 
reproduced by the method presented here. 
This will be investigated in more detail in
chapter~\ref{sec::chap3}. There it will be demonstrated that the high
energy expansion (including sufficiently many terms) and the Pad\'e
result agree remarkably well between $x=0$ and $x=0.7$ which covers most
of the perturbative region for charmed and bottom quarks.

The results which include high
moments up to $C_6$, $C_7$ or even $C_8$ are remarkably stable down to
very small values of $v$.
Different Pad\'e approximations of the same degree 
and approximants with a
reduced number of parameters give rise to practically identical
predictions, which could hardly be distinguished in 
Fig.~\ref{fighigh}. Minor variations are observed close
to threshold, {\it after} subtracting the singular and constant parts.
The remainder $\delta R$ for up to ten different Pad\'e 
approximants is shown in 
Fig.~\ref{figthr}. In~\cite{CheKueSte96} it was demonstrated that
there is a perfect agreement for $R^{(2)}_l$;
$R_{\it NA}^{(2)}$ seems to converge to the solid line 
($[4/4], [5/3]$ and $[3/5]$)
when more moments from small $q^2$ are included. The dashed lines are from
the $[3/3], [4/2], [2/4]$ and $[3/4]$, the dotted
ones from lower order Pad\'e approximants.
The dash-dotted curve is the [4/3] Pad\'e approximant and has a pole very 
close to $\omega=1 (1.07\ldots)$.
For the Abelian part a classification of the different results
can be seen: the dashed lines are $[4/2]$ and $[2/4]$,
the solid ones $[3/2], [2/3], [5/3], [3/5]$ and $[5/4]$ 
Pad\'e approximants.

Recently additional terms from the high energy expansion have been
injected in the Pad\'e approximation~\cite{rhdiss,CheHarSte98}.
\begin{figure}[tf]
 \begin{center}
 \begin{tabular}{cc}
   \small (a) & \small (b) \\[-2ex]
   \leavevmode
   \epsfxsize=6cm
   \epsffile[110 270 480 560]{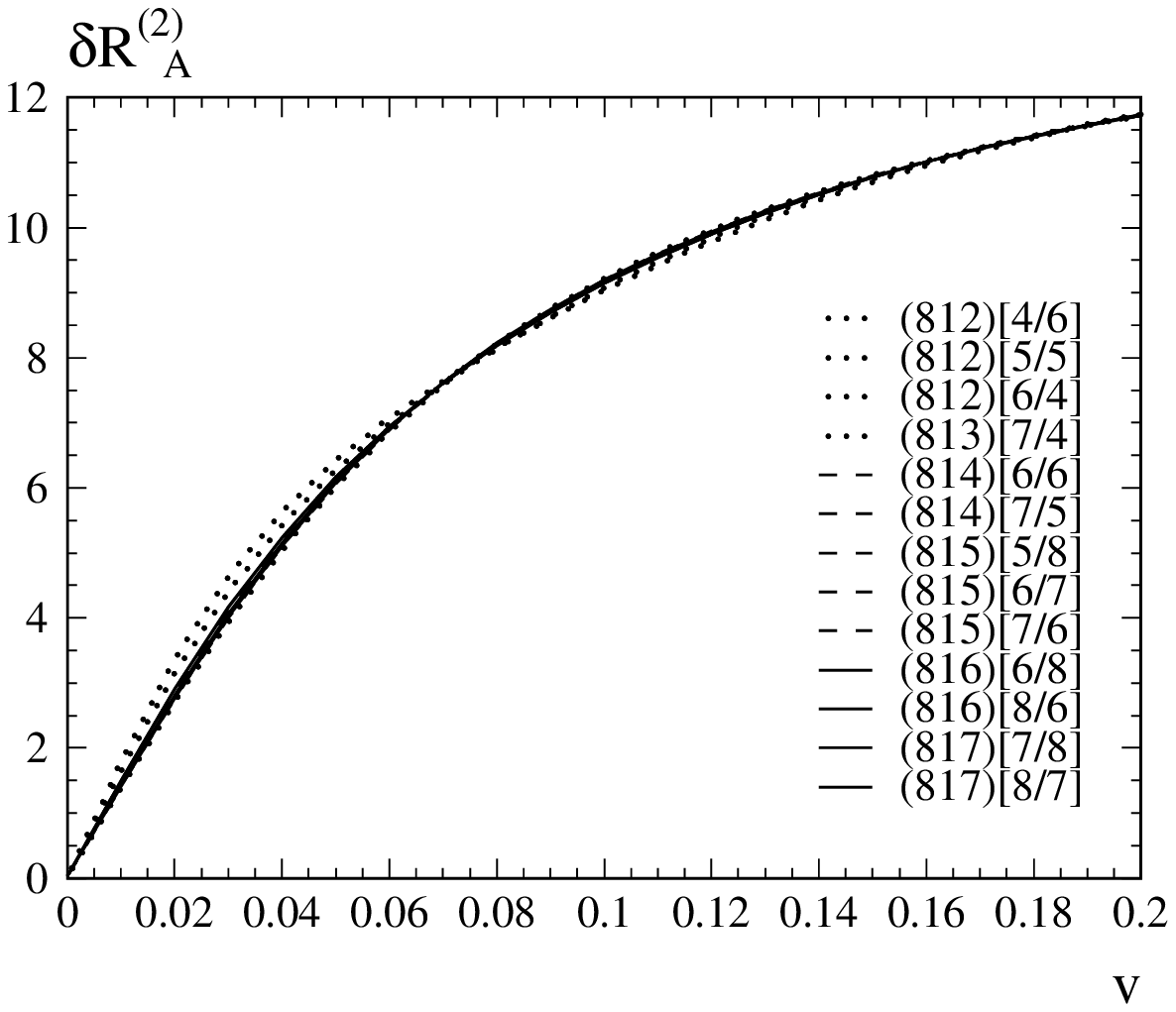 }
&
   \epsfxsize=6cm
   \epsffile[110 270 480 560]{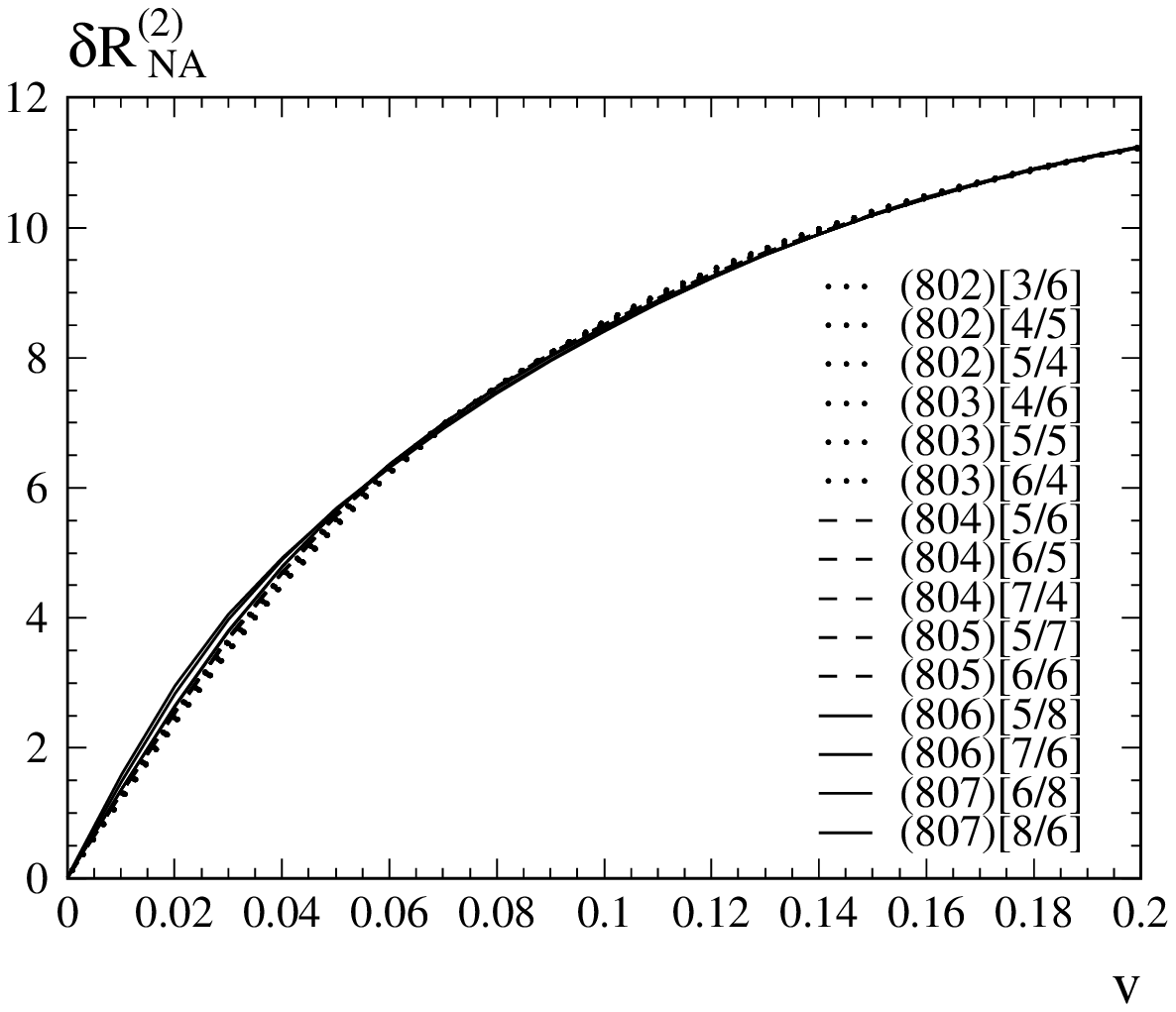 }
 \end{tabular}
   \caption[]{\label{fig::pademn}\sloppy
     Variation of the prediction for $R_A^{(2)}$ and $R_{NA}^{(2)}$ by
     including an increasing number of terms from the high energy
     expansion. The singular and constant pieces at $v=0$ have been
     subtracted. (From~\cite{rhdiss}.)  }
\end{center}
\end{figure}
The results, shown in Fig.~\ref{fig::pademn} confirm those from~\cite{CheKueSte96} and demonstrate the stability of the Pad\'e
approximation. It must be stressed that a safe estimate of the
remaining uncertainty in $R_A$ and $R_{NA}$ amounts to less than $0.02$
for $v$ above $0.1$ and is around $0.05$ in the region $v\approx 0.03$.
This region is, however, entirely dominated by the singular and constant
terms with values around 50 and higher. The perturbative predictions for
$R$ are therefore under excellent control.
It goes without saying that the function $\Pi(q^2)$ constructed this way
and evaluated e.g.\ in the Euclidean region could be a useful tool for
other investigations of interest, for example for sum rules involving
massive quarks.

Quite some effort has been invested in the analytic evaluation
of $R_A$ and $R_{NA}$ close to threshold. As stated above, the singular
and constant terms (Eqs.~(\ref{Ra}), (\ref{Rna}) and (\ref{Rnl})) were
derived through general considerations~\cite{BarGatKoeKun75,CheKueSte96}. 
To evaluate the $v\ln v$ and the $v$
terms, however, elaborate analytical calculations of the two-loop form
factor were used~\cite{hoang,CzaMel98} with the results
\begin{eqnarray}
R_A &=& 3\bigg\{ {\pi^4\over 8v} - 3\,\pi^2
                 +v\,\bigg(-{\pi^4\over 24}
\nonumber\\&&
                    +{3\over 2}\,
                    \big({\pi^4\over 6} 
+ \pi^2\,
                    (-{35\over 18} - {2\over 3}\,\ln v 
                    + {4\over 3}\,\ln 2) +
                         {39\over 4} - \zeta(3)\big)\bigg) \bigg\}
\\
R_{NA} &=& 3\bigg\{\pi^2\,\left({31\over 48} - {11\over 8}\,\ln 2v\right)
\nonumber\\&&
       +{3\over 2}\,v\,\bigg(\pi^2\,({179\over 72} - \ln v 
       - {8\over 3}\,\ln 2) 
       - {151\over 36} - {13\over 2}\,\zeta(3)\bigg)\bigg\}
\end{eqnarray}
    
\begin{figure}
\begin{center}
\begin{tabular}{lll}
    \leavevmode
    \epsfxsize=4.0cm
    \epsffile[110 265 465 560]{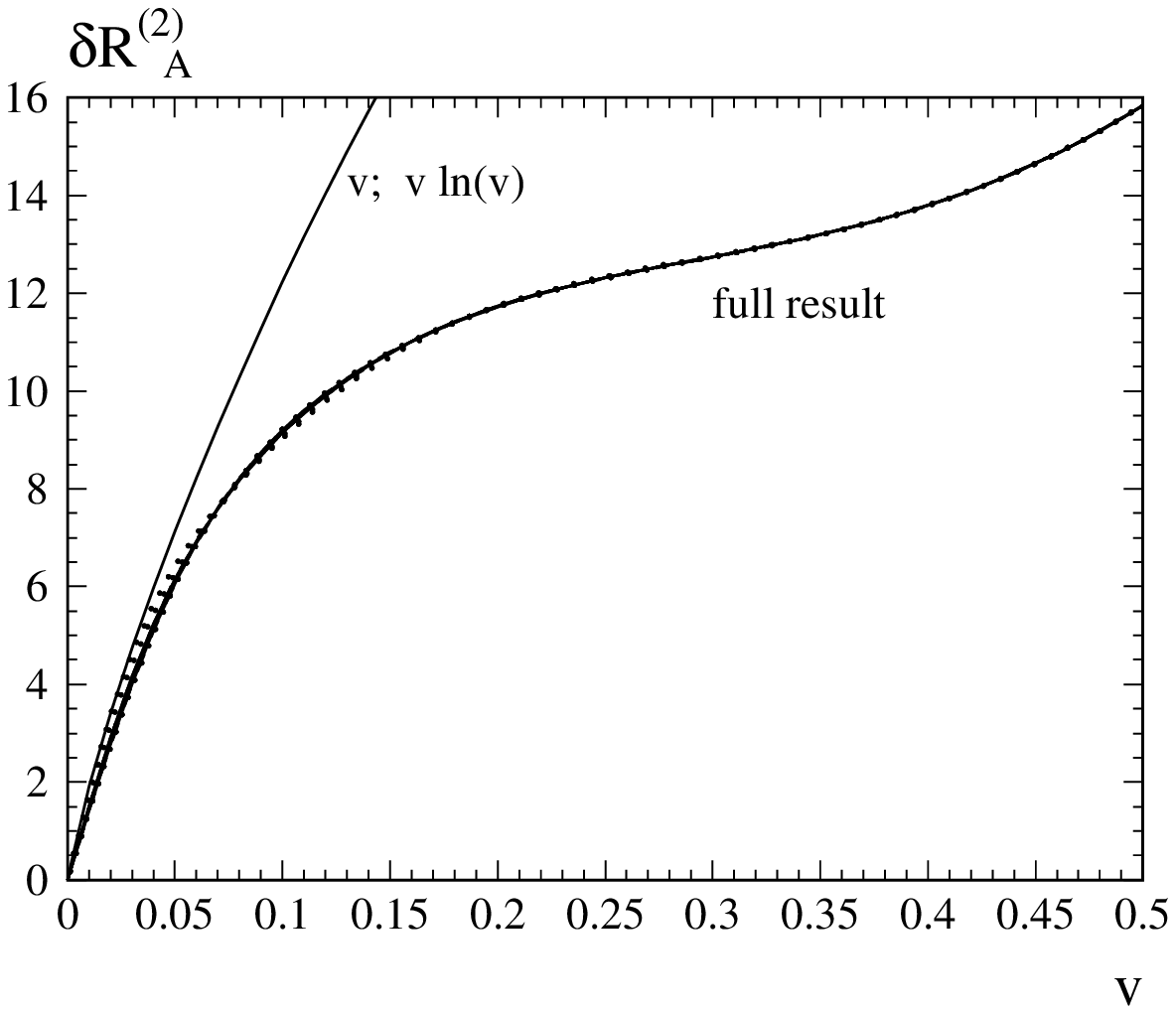} &
    \epsfxsize=4.0cm
    \epsffile[110 265 465 560]{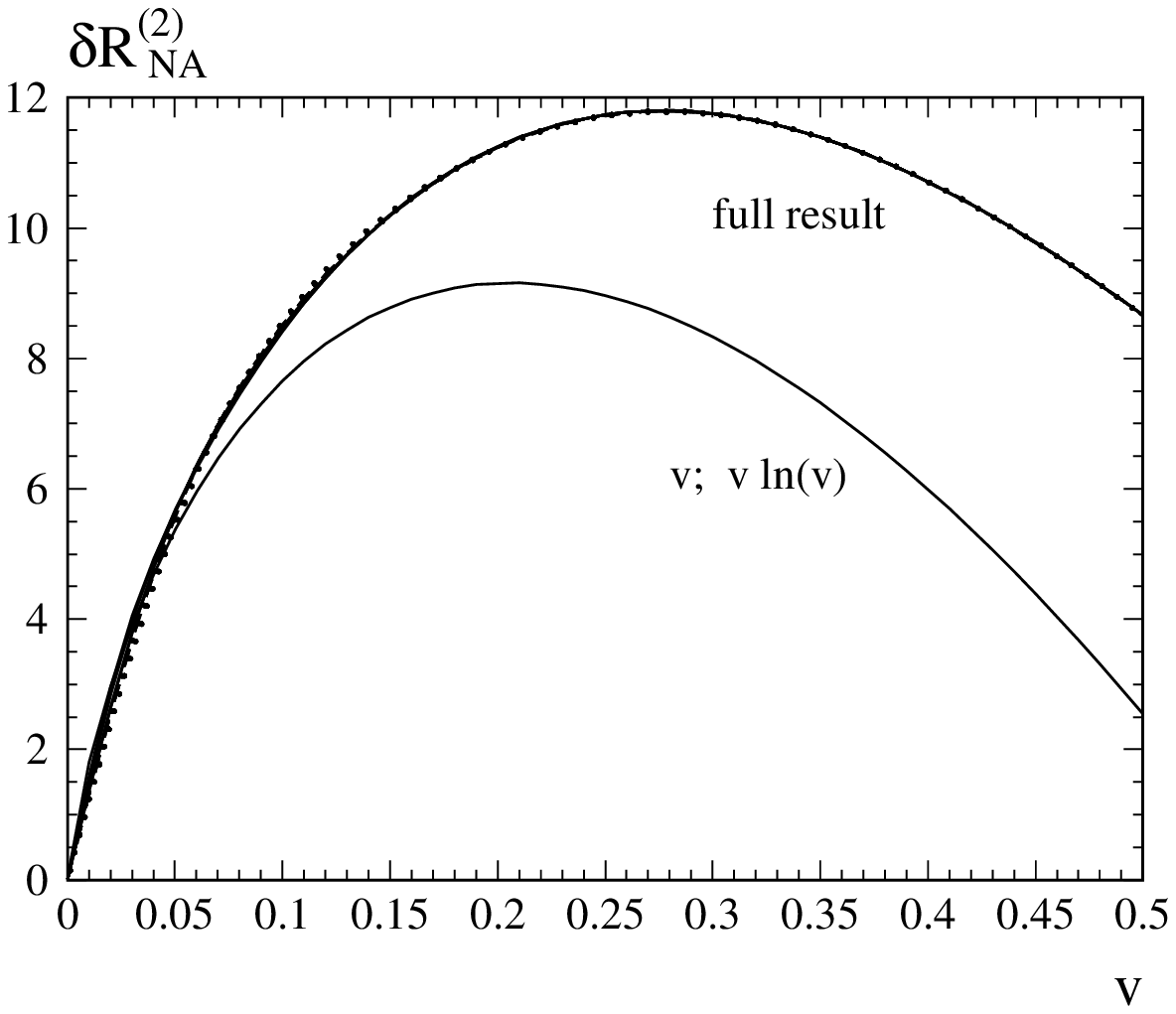} &
    \epsfxsize=4.0cm
    \epsffile[110 265 465 560]{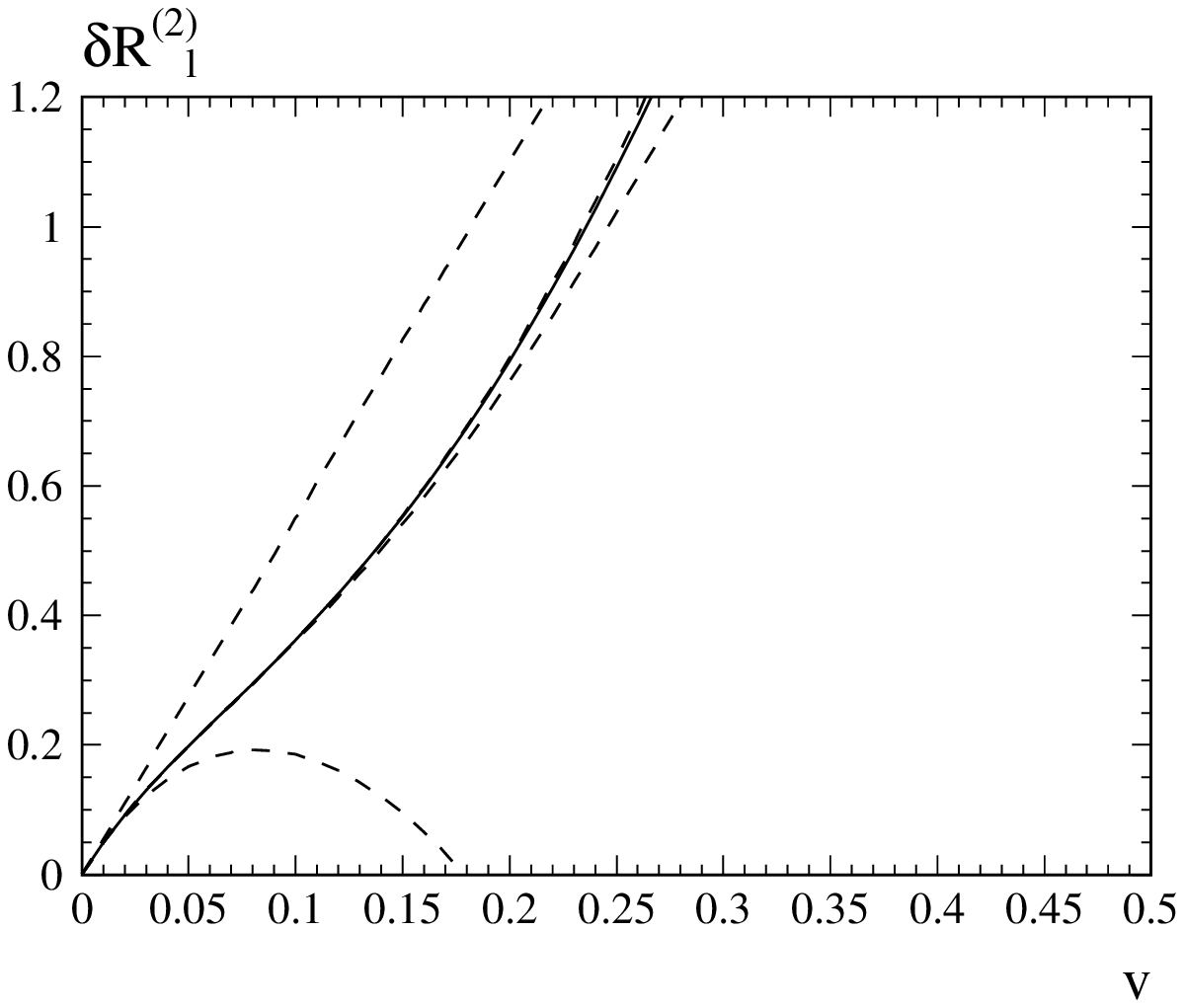}
\end{tabular}
\caption[]{\label{fig::rvlnv}\sloppy\rm
      Comparison of the small $v$ expansion (solid curves) of
      $\delta R_A$, $\delta R_{NA}$ with the full result (solid and dotted
      curves) after subtraction of the singular and constant pieces. The
      difference between dotted and solid curve indicate the remaining
      uncertainty of the semianalytical result. For $\delta R_l$
      successive approximations of the small $v$ expansion (dashed
      curves) and exact result (solid) are shown. }
\end{center}
\end{figure}

While the result for $R_A$ was still obtained in the framework of
``classical'' QED calculations, $R_{NA}$ was calculated using a
convenient technique which formalized the expansion for small
$v$ (see~\cite{BenSmi98,CzaMel98,BenSigSmi98}). 
Pad\'e and small $v$ results are compared in
Fig.~\ref{fig::rvlnv}, again subtracting first the singular and constant
pieces. The slopes for very small $v$ predicted by the two approaches
are again in nice agreement, giving further credibility to the Pad\'e
result. However, it is also clear from this comparison that the small
$v$ expansion alone cannot lead to a reliable prediction over a larger
energy range.  This is explicitely demonstrated for the case of $R_l$
where the analytic result is available. Terms at least of order $v^3$
are needed for a stable prediction.

\begin{figure}[t]
\begin{center}
\begin{tabular}{c}
\leavevmode
\epsfxsize=6cm
\epsffile[100 280 460 560]{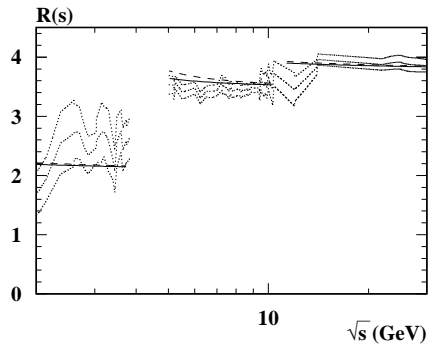}
\end{tabular}
\caption[]{\label{fig::rtot}
  $R(s)$ plotted against $\protect\sqrt{s}$.  The scale $\mu^2=s$ has
  been adopted. The dashed curves correspond to the values
  $M_c=1.8$~GeV, $M_b=5.0$~GeV and $\alpha_s(M_Z^2)=0.121$, whereas for
  the solid curves $M_c=1.4$~GeV, $M_b=4.4$~GeV and
  $\alpha_s(M_Z^2)=0.115$ is used.  The dotted lines show a recent
  compilation of the available experimental data. The central curves
  correspond to the mean values, upper and lower curves indicate the
  combined statistical and systematical errors. (From~\cite{chkst98}.)
  }
\end{center}
\end{figure}

A compilation of theoretical results can be found in~\cite{chkst98}
where the prediction for massive quark production is compared with the
measurements over a wide energy region (Fig.~\ref{fig::rtot}).

Up to this point only the vector current correlator has been discussed.
However, the techniques described above have also been applied to other
currents~\cite{CheKueSte96:2}: the axial vector, relevant e.g.\ for top
production through the virtual $Z$ boson, as well as scalar and
pseudoscalar currents describing for example the decay of Higgs bosons
into massive quarks. Recently also the singlet piece of top quark
production through the axial current was evaluated~\cite{HarSte97},
completing thus the $\order{\alpha_s^2}$ prediction of massive quark
production.

\section{Large Momentum Expansions\label{sec::chap3}}
An alternative route towards an efficient evaluation of the polarization
function is based on the large momentum
expansion.  
In this case the
polarization function $\Pi(q^2)$ is expanded in powers of $m^2/q^2$,
multiplied by logarithms of $m^2/q^2$, with the power of the logarithms,
however, limited by the number of loops under consideration. Technically
the expansion is given by a series of products of massless propagator
and massive tadpole integrals. 

\tabcolsep=0em

\begin{figure}
\begin{center}
\begin{tabular}{cc}
    \leavevmode
    \epsfxsize=5.5cm
    \epsfysize=3.5cm
    \epsffile[110 265 465 560]{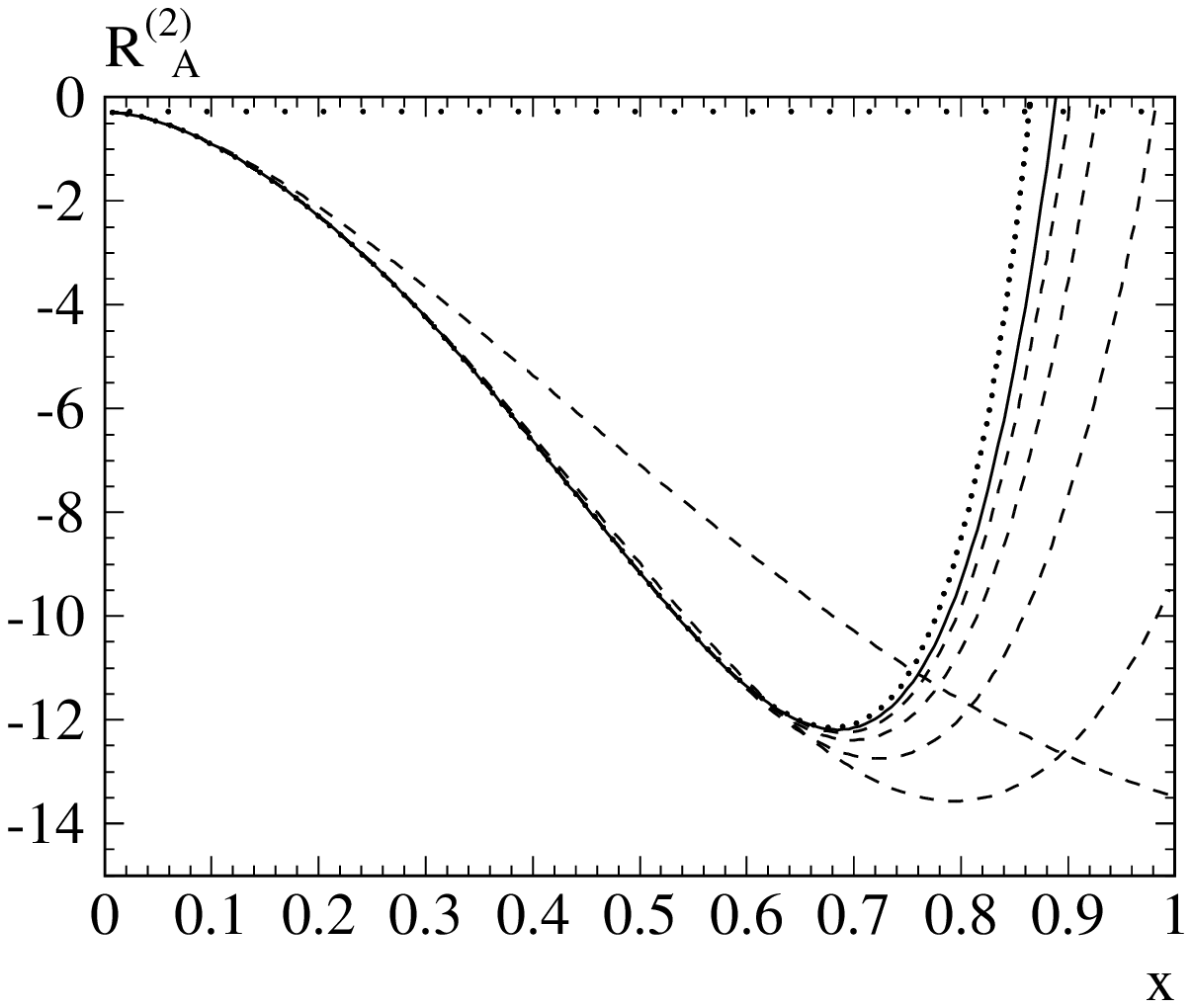}&
    \epsfxsize=5.5cm
    \epsfysize=3.5cm
    \epsffile[110 265 465 560]{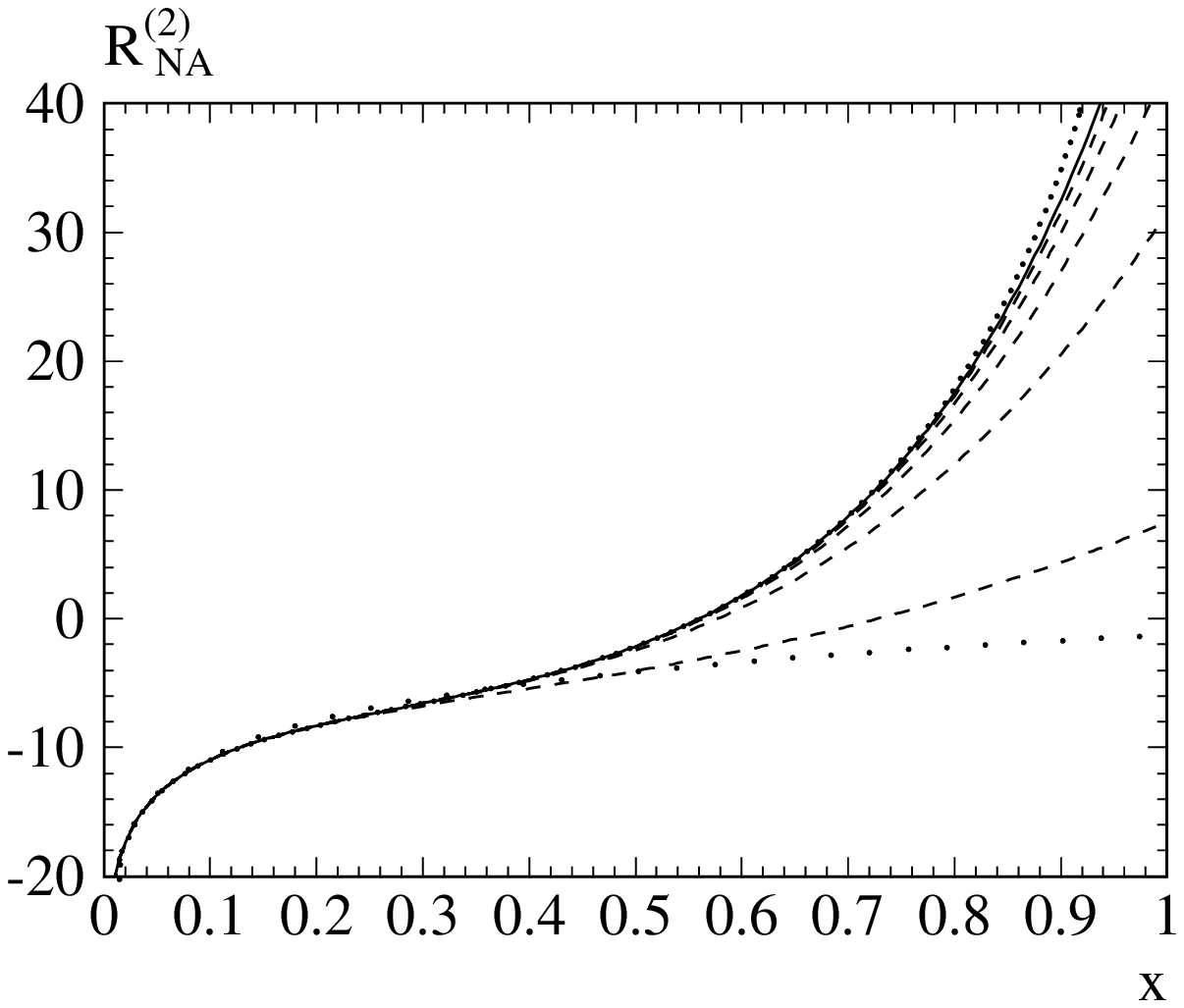}
\end{tabular}
    \caption[]{\label{rvx.ps}\sloppy\rm
      The Abelian contribution $R_A^{(2)}$ and the non-Abelian piece
      $R_{\it NA}^{(2)}$.  Wide dots: no mass terms; dashed lines:
      including successively more mass terms $(m^2/s)^n$ up to $n=5$;
      solid line: including mass terms up to $(m^2/s)^6$; narrow dots:
      semi-analytical result. The scale $\mu^2 = m^2$ has been adopted.
      (From~\cite{CheHarKueSte97}.)
      }
\end{center}
\end{figure}

In principle the full information on the analytic function is contained
in this series. The structure of the integrals is simplified by moving
from two- to one-scale integrals. However, an enormous proliferation of
the number of diagrams and the amount of algebraic calculation is
observed, requiring the development of programs which implement the
diagrammatic expansion, and translate the resulting diagrams into input
files for other programs which in turn are suited for the algebraic
evaluation of individual diagrams. One example for such a
``superprogram'' is {\tt GEFICOM}~\cite{geficom} which uses 
{\tt QGRAF}~\cite{qgraf} for the generation of diagrams, 
{\tt LMP}~\cite{rhdiss} or {\tt EXP}~\cite{Sei:dipl} for the diagrammatic
expansion through the hard mass or large momentum procedure, and {\tt
  MATAD}~\cite{Stediss} and {\tt MINCER}~\cite{mincer2} for the
evaluation of diagrams. Even nested expansions with a hierarchy of
several scales are possible in this framework. (A more detailed
description of the status of algebraic programs can be found in~\cite{HarSte:review}.)  After performing the expansion in $m^2/q^2$ up
to a given power, one may directly take the absorptive part and thus
predict $\Pi(q^2)$ in the high energy region. The comparison with the
Pad\'e result (discussed in chapter~\ref{sec::chap2}) shows excellent
agreement in the region of $x$ between zero and $0.7$ and thus down to
fairly low energies (Fig.~\ref{rvx.ps}).

One may even suspect that, given sufficiently many terms of the
absorptive part alone, an approximation of arbitrary precision can be
achieved. This is indeed true for the two-loop result and the ``double
bubble'' contribution with massless quarks in the internal loop, which
are available in analytical form and thus can be used as ``toy models''.
However, in both examples, the only threshold is located at $\sqrt{s} =
2m$ and convergence down to this point is naturally expected. In
contrast, the other diagrams have four-particle cuts at $\sqrt{s} = 4m$,
suggesting convergence only above this point.  This has been confirmed
by a detailed study~\cite{CheHarKueSte97} of the double bubble
contribution with equal masses in both fermion loops, where inclusion of
an increasing number of terms does not lead to an improvement beyond
$x\approx 0.5$.  Given both real and imaginary parts, this problem could
be and has been circumvented as discussed in chapter~\ref{sec::chap2}.
However, below we will be interested in the situation where the
absorptive piece only is available.

 Nevertheless, the relative smallness of the four-particle
contribution and the slow opening of the phase space reduce  this effect
and a fairly good approximation of $R_A$ and $R_{NA}$ is achieved even
for $x=0.7$, thus covering most of the interesting energy range (cf.\ 
Table~\ref{tab::1}).  The same technique has also been used for the
axial, the scalar and the pseudoscalar current~\cite{HarSte97,HarSte98}.
(This motivates the step to $\alpha_s^3m^4$ in chapter~\ref{sec::chap4}.)
The agreement of the two-loop result described in
chapter~\ref{sec::chap2} with the threshold and the large momentum
expansion in the respective ranges of validity demonstrate that the
perturbative NLO prediction of $R(s)$ for massive quarks is under full
control.

\section{Quartic Mass Terms in NNLO form Operator Product Expansion
  \label{sec::chap4}}
Fig.~\ref{rvx.ps} suggests that the first three terms provide an
excellent approximation at $x=0.5$ and are quite acceptable even at
$x=0.7$. Using this line of reasoning, a possible route for a NNLO
prediction (order $\alpha_s^3$) of $R(s)$, including quark mass effects,
is at hand. The massless result~\cite{GorKatLar91} and the $m^2/s$ terms~\cite{CheKue90} have been obtained a long time ago. The strategy used in~\cite{CheKue94} allowed to predict the $\alpha_s^2 m^4/s^2$ terms in
$R(s)$ by evaluating two-loop~(!) tadpoles and massless propagators
only. Additional ingredients are the operator product expansion and the
renormalization group equations, plus certain anomalous dimensions.
Using this method and algebraic programs it is thus possible to obtain
the $\alpha_s^3 m^4/s^2$ terms from three-loop tadpoles and massless
propagators~\cite{rhdiss,chkprep}. Let me briefly describe this method:
In a first step the OPE is applied to the time ordered product of two
currents
\begin{eqnarray}
\int \dd x e^{iqx} T\big(j_\mu(x)j_\nu(0)\big) &=& (q_\mu q_\nu -
g_{\mu\nu}q^2)\bigg\{ A(q^2,\mu^2,\alpha_s){\bf 1} +
\label{eq::tprod}
\\&&
B(q^2,\mu^2,\alpha_s) {\bar m^2\over q^2}(\mu^2) + \sum_{n=1}^6 {1\over
  q^4} C_n(q^2,\mu^2,\alpha_s){\cal O}_n\bigg\}\,.
\nonumber
\end{eqnarray}
Only three of the six operators ${\cal O}_n$ with dimension four are
gauge invariant and contribute to physical matrix elements:
\begin{equation}
{\cal O}_1 = G^2, \qquad {\cal O}_2 = m\bar q q, \qquad {\cal O}_6 =
\bar m^4(\mu^2)\,,
\end{equation}
the others are required for the proper construction to the coefficient
functions $C_n$. 
For the NNLO calculation, $C_1$, $C_2$ and $C_6$ are required up to
${\cal O}(\alpha_s^2)$, ${\cal O}(\alpha_s^3)$ and ${\cal
  O}(\alpha_s^2)$ respectively. To obtain these coefficient functions,
only massless propagator type integrals, at most of three loop, are
needed.
To calculate the vacuum matrix elements of ${\cal O}_1$ and ${\cal
  O}_2$, massive tadpole integrals -- at most of three loop -- are
needed. In a last step one uses renormalization group invariance of the
dimension four part of Eq.~(\ref{eq::tprod}). Employing the anomalous
dimension matrix~\cite{CheSpi88}  of ${\cal O}_{1,2,6}$ one finally obtains the
coefficients of the terms $\alpha_s^3 \bar m^4 \ln^n q^2/\mu^2$ with
$n=1,2,3$. Only these terms contribute to the absorptive part and one
finally arrives at~\cite{GorKatLar91,CheKue90,rhdiss,chkprep}
\begin{eqnarray}
  R^v(s) &=& 3\,\bigg\{1 - 6\,x^2 + {\alpha_s\over
    \pi}\,\bigg[1 + 12\,x - 22\,x^2\bigg] 
+ \left({\alpha_s\over \pi}\right)^2\,\bigg[1.40923 
  \nonumber\\&&\mbox{}
+  104.833\,x + 
  x^2\,(139.014 - 4.83333\,l_x)\,
  \bigg] 
+ \left({\alpha_s\over \pi}\right)^3\,\bigg[-12.7671 
  \nonumber\\&&\mbox{}
+  541.753\,x 
+ x^2\,(3523.81 - 158.311\,l_x +
  9.66667\,l_x^2)\,\bigg]\bigg\}\,,
\end{eqnarray}
with $x\equiv \bar m^2(s)/s$, $l_x \equiv \ln (\bar m^2(s)/s)$ and $n_f
= 5$.

\section{Expansion techniques and electroweak interactions\label{sec::chap5}}
Electroweak observables are frequently affected by the interplay between
strong and electroweak interactions. Important examples are the hadronic
contributions to the running of the QED coupling from the Thompson limit
to $M_Z$, QCD effects on the $\rho$ parameter and related quantities,
and ``mixed'' vertex corrections affecting for example the $Z$ decay
rate. Quark mass effects and their perturbative treatment are important
for $\alpha_{\rm QED}(M_Z)$. A detailed discussion of the last topic is
beyond the scope of this paper and can be found in~\cite{chkst98,KueSte98}. Let us present some aspects of the two
remaining items.

{\it Gauge boson self-energies, the mass of the top quark and QCD:}
The indirect determination of the top quark through quantum corrections
prior to its observation in hadronic collisions can be considered one of
the triumphs of the Standard Model. The experimental precision of the
key observables, the masses of the top quark and the $W$ boson together
with the weak
mixing angle as determined by asymmetry measurements has increased
during the past years and this process will continue in the foreseeable
future. In order to control the influence of the top quark at an
adequate level the inclusion of QCD corrections in the top and bottom
induced self energies is mandatory. The dominant terms are characterized
by the $\rho$ parameter which, in lowest order, is given by~\cite{veltman}
\begin{equation}
\Delta\rho = 3 \sqrt{2} \frac{G_F m_t^2}{16 \pi^2}.
\end{equation}
In view of the large difference between pole and running mass at scale
$m_t$
\begin{equation} 
m_t(pole)=\bar m_t(m_t) \left(1+\frac{4}{3}\frac{\alpha_s}{\pi}+\cdots\right)
\end{equation} 
inclusion of two and even three-loop QCD corrections to $\Delta\rho$ is
mandatory. 

Analytic results are available in two-loop approximation not only for
the leading term~\cite{djouadi1} in $\Delta\rho$ but also for all self
energies, with arbitrary top and bottom masses~\cite{Kniehl}. The
resulting shift in the prediction for $M_W$ for fixed $G_F$,
$\alpha_{\rm QED}$, $M_Z$ and $m_t=175$~GeV amounts to 68 MeV, well
comparable to the present precision and significantly larger than the
anticipated accuracy of roughly 30 MeV. To arrive at a precise
prediction for the central value and to control the theoretical
uncertainty, three-loop QCD contributions to $\Delta\rho$ as well as to
$\Delta r$ are required. The $\rho$ parameter~\cite{rhoparam} can again
be expressed through diagrams with vanishing external momentum (vacuum
or tadpole diagrams), the remaining quantities involve two point
functions at non-vanishing external momentum and can be
obtained~\cite{deltar} through an expansion in the small mass ratio
$M_Z^2/m_t^2$. The leading three-loop term corresponds to a shift of
$-10.9$~MeV.  The first three terms, amounting to $-13.7$~MeV, are adequate
for a prediction with an accuracy better than 1 MeV (Table~\ref{tab2}).
Conversely, this combined shift is equivalent to a reduction of the
effective top mass by about 1.6 GeV.
\tabcolsep=.2em
\begin{table}[th]
\begin{center}
\begin{tabular}{|l||r|r|r|}
\hline
$\delta M_W$ in MeV  & $\alpha_s^0$ & $\alpha_s^1$ & $\alpha_s^2$ \\
\hline
\hline
$M_t^2$             & 611.9 & -61.3 & -10.9 \\
const.              & 136.6 & -6.0  & -2.6  \\
$1/M_t^2$           & -9.0  & -1.0  & -0.2  \\
\hline
\end{tabular}
\end{center}
\caption[]{\label{tab2} The change in $M_W$ separated according
                      to powers in $\alpha_s$ and $M_t$ in the on-shell
                      scheme. (From~\cite{deltar}.)}
\end{table}
\tabcolsep=0em

To exploit the precision expected from a future linear collider which
will pin down $m_t$ to better than 200 or perhaps even 100 MeV, the
inclusion of $\alpha_s^2$ terms is evidently mandatory.

Similar considerations~\cite{deltar} are valid for the effective
weak mixing angle which can be deduced from the left right asymmetry or
the forward backward asymmetry in a straightforward way.

An important issue in this connection is the size of uncertainties,
arising either from uncalculated higher orders, or from the
``parametric'' uncertainties in $\alpha_s$ and $m_t$. Shifts of
$\delta\alpha_s = 0.003$ and $\delta m_t = 5$~GeV lead to changes 
in $M_W$ of $-2.4$~MeV and 35~MeV respectively. The uncertainty from 
uncalculated higher orders is
completely negligible. A $W$-mass determination with a
precision in the 10~MeV range should therefore be accompanied by a top
mass measurement with a precision around or better than 1~GeV.

{\it Mixed QCD and electroweak vertex corrections:}
As stated above, gauge boson self energies, in particular those induced
by fermion loops, give rise to the dominant radiative corrections for
electroweak precision observables. Nevertheless, for a complete
treatment of ${\cal O}(\alpha\alpha_s)$ the inclusion of irreducible
vertex corrections is necessary. These have to be distinguished from the
reducible ones which are easily incorporated, if the electroweak result,
including one-loop weak terms, is simply multiplied by the QCD
correction factor $(1+\alpha_s/\pi + \cdots)$. 

\begin{figure}[t]
\leavevmode
\centering
\epsfxsize=8cm
\epsffile[105 558 503 717]{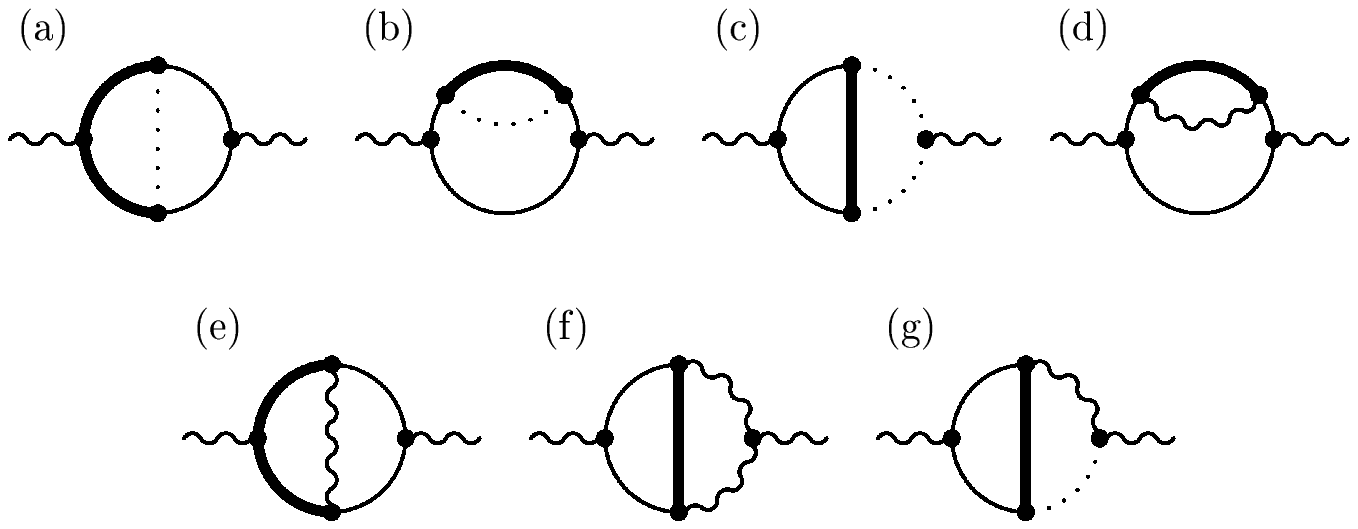}
\caption[]{Diagrams contributing to $\delta\Gamma^W_b$
  in ${\cal O}(\alpha)$. Thin lines correspond to bottom quarks, thick
  lines to top quarks, dotted lines to Goldstone bosons and inner wavy
  lines represent $W$ bosons.
  \label{fig::2l}
  }
\end{figure}

The physics and the techniques of calculation are markedly different for
vertices leading to light ($u$, $d$, $s$ and $c$) quark pairs on one
hand~\cite{CzarK} and for decays into $b\bar b$ on the other
hand~\cite{fjrt}, a consequence of the presence of top quarks with their
enhanced contribution to the vertex proportional $m_t^2/M_W^2$. The
irreducible one-loop vertex diagrams are shown in Fig.~\ref{fig::2l}.

To obtain all one-particle irreducible vertex diagrams in two-loop
approximation , these have to be dressed with gluons in all conceivable
ways. The resulting amplitudes are first studied for arbitrary $q^2$ by
considering expansions in the ratio $x_Z=q^2/M_Z^2$ and $x_W=
q^2/m_W^2$ (or for some diagrams, in $1/x_{Z,W}$). Even for the limiting
values $x_Z=1$ and $x_W= M_Z^2/M_W^2 $ the exact results are well
approximated, once sufficiently many terms are included. Using $\alpha_s
=0.12$, $\alpha=1/129$, $\sin\theta_W = 0.223$ $M_Z = 91.19$~GeV, it is
found~\cite{CzarK} that the net effect of the nonfactorizable corrections
is
\begin{eqnarray}
\lefteqn{\Gamma^{\mbox{\small (2-loop EW/QCD)}} - {\alpha_s\over \pi}
\Gamma^{\mbox{\small (1-loop EW)}}}
\nonumber \\
&& \qquad\qquad = 
\left\{
\begin{array}{l}
-1.13(4)\times 10^{-4}  \mbox{ GeV}\quad \mbox{for $Z\to \bar uu$}
\\
 -1.60(6)\times 10^{-4}  \mbox{ GeV}\quad \mbox{for $Z\to \bar dd$}
\end{array}
\right.
\end{eqnarray}
The total change in the partial width $\Gamma(Z\to hadrons)$ is
obtained by summing over 2 down-type and 2 up-type quarks:
\begin{eqnarray}
\Delta \Gamma(Z\to u,d,s,c)= -0.55(3) \mbox{ MeV}
\end{eqnarray}
which translates into the change of the strong coupling constant
determined at LEP 1 equal to
\begin{eqnarray}
\Delta\alpha_s = 
-\pi{\Delta \Gamma(Z\to hadrons)\over \Gamma(Z\to hadrons) }
=\pi{0.55\over 1741} \approx 0.001
\end{eqnarray}
This shift is somewhat smaller but still of the same order of
magnitude as the 
experimental accuracy and should
to be taken into account in the  final analysis of LEP 1 data.
Electroweak parameters extracted from $Z$ decays are not affected
by this correction.

The two-loop corrections to the $Zb\bar b$ vertex are dominated by terms
quadratic in $m_t$. The QCD corrections to these have been evaluated
already some time ago~\cite{fjrt}.  The sub-leading terms $\propto \ln
m_t^2$ were obtained in~\cite{KwiSte95} and are of comparable magnitude,
thus indicating relatively slow convergence of the series. The complete
evaluation is thus mandatory and has been recently been
performed~\cite{HarSeiSte97}. In contrast to the previous case one is
confronted with three different scales, the masses of top, the $W$ and
the $Z$ boson.

Using the \hmp\ for $m_t^2\gg M_Z^2,M_W^2$, one may factor out the
$m_t$--dependence. However, for a part of the diagrams one still is left
with two-scale and even three-scale integrals involving $M_Z^2$ and
$M_W^2$ and $\xi_W M_W^2$, where $\xi_W$ is the electroweak gauge
parameter which has been kept in~\cite{HarSeiSte97}.  Although they
appear to be only one-loop integrals, their exact evaluation up to
${\cal O}(\epsilon)$ is inconvenient.  Instead, the
results~\cite{HarSeiSte97} were obtained by applying the \hmp\ to these
kinds of diagrams once more, this time using $\xi_W M_W^2, M_W^2\gg
M_Z^2$. This seemingly unrealistic choice of scales can be well
justified: It is not possible for an expansion to distinguish the
inequality $M_W^2\gg M_Z^2$ from $4M_W^2\gg M_Z^2$ or $(m_t+M_W)^2\gg
M_Z^2$, the latter ones being perfectly alright. The only matter is to
perform the expansion on the appropriate side of all thresholds, and
here one is concerned with thresholds at $2M_W$ and at $m_t+M_W$.
Therefore, the choice $M_W^2\gg M_Z^2$ is to be understood purely in
this technical sense. Graphically this continued expansion looks as
follows:
\begin{center}
\leavevmode
\epsfxsize=10cm
\epsffile[145 630 550 675]{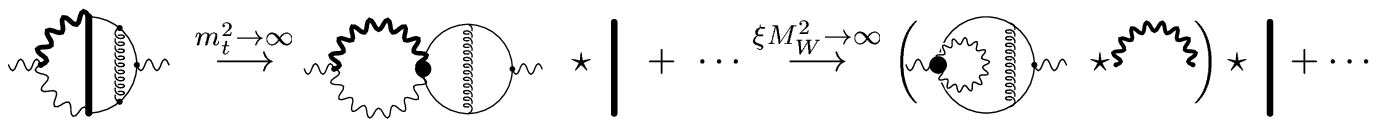}
\end{center}
where only those terms are displayed which are relevant in the
discussion above and all others contributing to the \hmp\ are merged
into the ellipse.  The thick plain line is the top quark, the thick wavy one
a Goldstone boson with mass squared $\xi_W M_W^2$, for example.  The thin
plain lines are $b$--quarks, the inner thin wavy lines are $W$--bosons,
the outer ones $Z$-bosons. The spring-line is a gluon. The mass
hierarchy is assumed to be $m_t^2\gg \xi_W M_W^2 \gg M_W^2 \gg
M_Z^2$. The freedom in choosing the magnitude of $\xi_W$ provides a
welcome check of the routines and the results.

The outcome of this procedure is a nested series: The coefficients of
the $M_W/m_t$--expansion are in turn series in $M_Z/M_W$. Note that in
contrast to the decay into $u,d,s,c$ there is no threshold at $M_W$
which makes an additional expansion in $M_W/M_Z$ unnecessary.

In view of this calculation the procedure of successive application of
the \hmp\ resp.\ the \lmp\ has been implemented in a Fortran 90 program
named {\tt EXP}~\cite{Sei:dipl}. Therefore, given an arbitrary
hierarchy of mass scales, the computation of a three-loop two-point
function can now be done fully automatically. Even more, the link to the
Feynman diagram generator {\tt QGRAF}~\cite{qgraf} in a common
environment called {\tt GEFICOM}~\cite{geficom} allows to obtain the
result of a whole physical process without any human interference except
for specification of the process and final renormalization.

Finally, the result for the $W$--induced corrections to the $Z$--decay
rate $\delta\Gamma^W(Z\to b\bar b)$ is conveniently presented in the
form of the renormalization scheme independent difference to the decay
rate into $d\bar d$. Inserting the on-shell top mass $m_t = 175$~GeV,
the $Z$--mass $M_Z=91.19$~GeV and $\sin^2\theta_W = 0.223$
gives~\cite{HarSeiSte97}
\begin{eqnarray}
&&\delta\Gamma^W(Z\to b\bar b) - \delta\Gamma^W(Z\to d\bar d) =
    \Gamma^0 {1\over \sin^2\theta_W} {\alpha\over \pi}
  \bigg\{ - 0.50  + (0.71 -0.48)
\nonumber\\&&\mbox{\hspace{0em}}
  + (0.08 - 0.29) + (-0.01 - 0.07) + (-0.007 - 0.006) 
  + {\alpha_s\over \pi} \bigg[ 1.16 
+ (1.21 
\nonumber\\&&\mbox{\hspace{0em}}
- 0.49) 
+ (0.30 - 0.65)
  +  (0.02 - 0.21 + 0.01) + (-0.01 - 0.04 + 0.004) \bigg] \bigg\}
\nonumber\\&&\mbox{\hspace{0em}}
=\Gamma^0 {1\over \sin^2\theta_W} {\alpha\over \pi} \bigg\{- 0.50 - 0.07 +
{\alpha_s\over \pi} \bigg[ 1.16 + 0.13 \bigg]\bigg\}\,,
\label{eqgamnum}
\end{eqnarray}
where the factor $\Gamma^0\alpha/(\pi\sin^2\theta_W)$ with $\Gamma^0 =
M_Z \alpha/(4 \sin^2\theta_W \cos^2\theta_W)$ has been pulled out for
convenience. The numbers after the first equality sign correspond to
successively increasing orders in $1/m_t^2$, where the brackets collect
the corresponding constant, $\log m_t$ and, if present, $\log^2
m_t$--terms. The numbers after the second equality sign represent the
leading $m_t^2$--term and the sum of the sub-leading ones.  The ${\cal
  O}(\alpha)$ and ${\cal O}(\alpha\alpha_s)$--results are displayed
separately. Comparison of this expansion of the one-loop terms to the
exact result~\cite{AkhBarRie86} shows agreement up to $0.01\%$
which gives quite some confidence in the $\alpha\alpha_s$--contribution.
One can see that although the $m_t^2$--, $m_t^0$-- and $m_t^0\log
m_t$--terms are of the same order of magnitude, the final result is
surprisingly well represented by the $m_t^2$--term, since the sub-leading
terms largely cancel among each other.

The uncertainty from uncalculated higher order QCD terms is far below
the foreseeable experimental precision, and the parametric uncertainty
from $\alpha_s$ and $m_t$ dominates. An important lesson to be learned
from the $Z b\bar b$ vertex in one- and two-loop approximation concerns
the interplay between ``dominant'' and ``sub-leading'' pieces: whenever
leading and sub-leading terms are of comparable magnitude, inclusion of
a sizeable number of terms in the expansion is required. The estimate of
the final result or of theoretical uncertainties based on the first two
terms of the series may lead to a wrong or misleading result.

\section{Summary\label{sec::chap6}}
Expansion techniques for Feynman amplitudes combined with sophisticated
computer algebra programs have lead to remarkable progress in multiloop
calculations during the past years. Problems with different mass and
energy scales can be treated with nested series. Powerful computers
allow to evaluate many terms in these expansions, and smallness of the
expansion parameter is thus no longer required. These techniques have
been successfully applied to purely hadronic as well as to electroweak
observables.
\vspace{1em}

{\bf Acknowledgments:} I would like to thank Joan Sola for organizing
this pleasant and very successful conference. The material presented in
this review has been developed in enjoyable and fruitful collaborations
with K.~Chetyrkin, A.~Czarnecki, R.~Harlander, Th.~Seidensticker and
M.~Steinhauser. The paper would never have been completed without the
\TeX nical help of R.~Harlander.  Work supported by {\it
  DFG-Forschergruppe ``Quantenfeldtheorie, Computeralgebra und
  Monte-Carlo-Simulationen''} (DFG Contract KU~502/6-1) and BMBF
Contract 056~KA~93~P6 at the University of Karlsruhe.

\def\app#1#2#3{{\it Act.~Phys.~Pol.~}{\bf B #1} (#2) #3}
\def\apa#1#2#3{{\it Act.~Phys.~Austr.~}{\bf#1} (#2) #3}
\def\cmp#1#2#3{{\it Comm.~Math.~Phys.~}{\bf #1} (#2) #3}
\def\cpc#1#2#3{{\it Comp.~Phys.~Commun.~}{\bf #1} (#2) #3}
\def\epjc#1#2#3{{\it Eur.\ Phys.\ J.\ }{\bf C #1} (#2) #3}
\def\fortp#1#2#3{{\it Fortschr.~Phys.~}{\bf#1} (#2) #3}
\def\ijmpa#1#2#3{{\it Int.~J.~Mod.~Phys.~}{\bf A #1} (#2) #3}
\def\jcp#1#2#3{{\it J.~Comp.~Phys.~}{\bf #1} (#2) #3}
\def\jetp#1#2#3{{\it JETP~Lett.~}{\bf #1} (#2) #3}
\def\mpl#1#2#3{{\it Mod.~Phys.~Lett.~}{\bf A #1} (#2) #3}
\def\nima#1#2#3{{\it Nucl.~Inst.~Meth.~}{\bf A #1} (#2) #3}
\def\npb#1#2#3{{\it Nucl.~Phys.~}{\bf B #1} (#2) #3}
\def\nca#1#2#3{{\it Nuovo~Cim.~}{\bf #1A} (#2) #3}
\def\plb#1#2#3{{\it Phys.~Lett.~}{\bf B #1} (#2) #3}
\def\prc#1#2#3{{\it Phys.~Reports }{\bf #1} (#2) #3}
\def\prd#1#2#3{{\it Phys.~Rev.~}{\bf D #1} (#2) #3}
\def\pR#1#2#3{{\it Phys.~Rev.~}{\bf #1} (#2) #3}
\def\prl#1#2#3{{\it Phys.~Rev.~Lett.~}{\bf #1} (#2) #3}
\def\pr#1#2#3{{\it Phys.~Reports }{\bf #1} (#2) #3}
\def\ptp#1#2#3{{\it Prog.~Theor.~Phys.~}{\bf #1} (#2) #3}
\def\sovnp#1#2#3{{\it Sov.~J.~Nucl.~Phys.~}{\bf #1} (#2) #3}
\def\tmf#1#2#3{{\it Teor.~Mat.~Fiz.~}{\bf #1} (#2) #3}
\def\tmph#1#2#3{{\it Theor.~Math.~Phys.}{\bf #1} (#2) #3}
\def\yadfiz#1#2#3{{\it Yad.~Fiz.~}{\bf #1} (#2) #3}
\def\zpc#1#2#3{{\it Z.~Phys.~}{\bf C #1} (#2) #3}
\def\ibid#1#2#3{{ibid.~}{\bf #1} (#2) #3}

\sloppy
\raggedright

\end{document}